# Understanding the anisotropic response of $\beta$-Ga$_2$O$_3$ to ion implantation

D. M. Esteves[1,2,*], R. He[3], S. Magalhães[2,4], M. C. Sequeira[5], Â. R. G. Costa[6], J. Zanoni[7], J. Rodrigues[7], T. Monteiro[7], F. Djurabekova[3], K. Lorenz[1,2,4], M. Peres[1,2,4]

[1] INESC Microsystems and Nanotechnology, Rua Alves Redol 9, Lisboa 1000-029, Portugal

[2] Institute for Plasmas and Nuclear Fusion, Instituto Superior Técnico, University of Lisbon, Av. Rovisco Pais 1, Lisboa 1049-001, Portugal

[3] Department of Physics, University of Helsinki, P.O. Box 43, FI-00014, Helsinki, Finland

[4] Department of Nuclear Science and Engineering, Instituto Superior Técnico, University of Lisbon, Estrada Nacional 10, km 139.7, Bobadela 2695-066, Portugal

[5] Institute of Ion Beam Physics and Materials Research, Helmholtz-Zentrum Dresden-Rossendorf, Bautzner Landstraße 400, 01328 Dresden, Germany

[6] Centre for Nuclear Sciences and Technologies, Instituto Superior Técnico, University of Lisbon, Estrada Nacional 10, km 139.7, Bobadela 2695-066, Portugal

[7] i3N, Department of Physics, University of Aveiro, Campus Universitário de Santiago, Aveiro 3810-193, Portugal

* Corresponding author: duarte.esteves@tecnico.ulisboa.pt

**ABSTRACT:**

While $\beta$-Ga$_2$O$_3$ is considered a promising wide bandgap semiconductor, the impact of ion-induced defect formation and anisotropic elasticity remains poorly understood. Here, we combine a simulation and experiment X-ray diffraction (XRD) study of the strain-stress dynamics induced by ion implantation into $\beta$-Ga$_2$O$_3$ single-crystals with different surface orientations. The strain accumulation in the out-of-plane direction is observed by XRD to occur in an anisotropic manner, with compressive strain along the [010] direction and tensile strain along the directions perpendicular to (100) and (001). An anisotropic stress/strain accumulation model is proposed and probed via Molecular Dynamics (MD), showing an excellent agreement with the experiments. For higher damage levels, pole figures obtained both experimentally and by MD via a novel reciprocal-space projection method reveal an orientation-independent $\beta$-to-$\gamma$ phase transition, with a fixed crystallographic relationship between the polymorphs. By exploring the strain-stress dynamics in anisotropic systems, this work establishes a method to directly compare macroscale diffraction experiments and atomistic simulations and opens a new path to engineer the properties of such systems utilizing their anisotropic response to ion implantation/irradiation.

1. Introduction

$\beta$-Ga$_2$O$_3$ is one of the most promising ultra-wide bandgap semiconductors for future (opto)electronic applications. Its bandgap of ∼4.9 eV at room temperature and the large breakdown electric field of ∼8 MV/cm lead to a Baliga figure of merit larger than that of other wide bandgap semiconductors such as GaN or SiC[1], thus driving applications in the realms of high-power electronics[2], optoelectronic devices[3], in particular solar-blind photodetectors[4–6], among others.

Ion implantation is a routine technique in the silicon industry, but it is still poorly understood in emerging wide bandgap semiconductors, in particular in Ga$_2$O$_3$. Nevertheless, promising results regarding controlled doping[7] and device processing underline the potential of ion implantation, especially for the design of vertical devices[8]. Fundamental studies further revealed intriguing responses upon ion irradiation and implantation, such as a disorder-induced phase transition to its defective spinel $\gamma$-phase[9,10], which occurs at about 0.78 displacements per atom (dpa)[11]. Moreover, the monoclinic system of $\beta$-Ga$_2$O$_3$ is particularly complex from the structural point of view, leading to anisotropic damage and strain, as observed when implanting Ga$_2$O$_3$ with different surface orientations[12,13]. Additionally, the two easy-cleavage planes (100) and (001)[14] trigger interesting effects upon ion implantation, such as the delamination and self-rolling of its surface layer, creating microtubes[11,15]. Hence, the physical mechanisms underlying the effects of ion implantation in this material are complex, and their analysis requires a synergistic approach between theory, simulation and experiments.

In this context, Molecular Dynamics (MD) is one of the key techniques for the atomistic simulation of materials. MD provides the possibility of assessing different types of phenomena at an atomistic scale and at very short time scales, some of which inaccessible experimentally[16]. For example, in the particular case of ion implantation in Ga$_2$O$_3$, a recently developed machine learning potential allowed to successfully reproduce the experimentally observed phase transition from the $\beta$- to the $\gamma$-phase[11,17]. On the flip side, it is often hard to directly compare MD results with experiments due to this scale mismatch. From the experimental point of view, there is a number of techniques relying on probing the reciprocal space of crystalline samples, including those based on X-ray and electron diffraction (XRD/ED)[18]. However, reciprocal-space approaches are often advantageous over real-space ones because they adequately capture long-range correlations[19] and can be more directly compared to experiments. For example, A. Boulle et al. have shown that it is possible to simulate the reciprocal space from MD simulations to calculate the rotation angle between crystalline domains and the microstrain tensor of irradiated materials[19]. In the case of epitaxial heterolayers, these methods yield information about the structural properties, namely phases, lattice constants, stress/strain states, tilts, twists, relative orientations, among other properties[20]. Despite significant efforts to study the structural properties via reciprocal space techniques based on computational methods, such as generative models[21], the direct comparison of diffraction phenomena between experiments and MD simulations is still underexplored[22].

In this work, a systematic study of ion implantation in $\beta$-Ga$_2$O$_3$ combining XRD experiments and MD simulations allowed to establish an atomistic model relating the effects of the collision cascades on the microscopic structure, as well as the anisotropic elastic properties of $\beta$-Ga$_2$O$_3$, to the macroscopic effects of ion implantation reported in the literature. Furthermore, the developed novel methodology for a direct comparison between experimental XRD pole figures and electron diffraction patterns with the appropriate projections of the reciprocal space, as calculated from MD simulations, can be applied to many other materials and processing techniques.

## 2. Results and discussion

### 2.1. Strain accumulation in the (100), (010) and (001) surfaces via X-ray diffraction

In order to assess the strain evolution with implantation fluence, three sets of samples with different surface planes, (100), (010) and (001), were implanted with 250 keV Cr ions with fluences up to $2.0 \times 10^{14}$ cm$^{-2}$. The symmetric high-resolution XRD (HRXRD) patterns were obtained for each sample, respectively, about the 400, 020 and 004 reflections. In order to extract the out-of-plane strain profiles, the diffractograms were fitted using the dynamic theory of XRD, as implemented in the multiple-reflection optimization package for XRD (MROX) software[23]. This software allows a diffractogram to be simulated from a layered structure, where the thickness, strain, $\varepsilon_\perp$, and crystalline quality (quantified by the Debye-Waller factor, DW) of each layer can be adjusted. The experimental $2\vartheta$–$\omega$ scans, as well as the fit results, are shown in Fig. 1 for the (010)- and (001)-oriented samples, while the HRXRD results for the (100)-oriented samples can be found elsewhere[15].

From this figure, it is already possible to notice the different responses to implantation for different surface orientations. Apart from the single, well-defined peak present on the pristine samples, ion implantation leads to the appearance of a second peak, as well as interference fringes known as Pendellösungen, which are very sensitive to the thicknesses of the layers[24]. The additional peak appears on the higher-$2\vartheta$ side for the (010)-oriented sample, but on the lower-$2\vartheta$ side for the (001)-oriented sample, which is in agreement with previous report on either orientation[25]. In light of Bragg's law, $2d \sin \theta = n\lambda$, where $\vartheta$ is the diffraction angle, $d$ is the interplanar spacing, $\lambda$ is the wavelength of the impinging X-rays and $n$ is an integer, this clearly suggests a contraction of the out-of-plane lattice parameter $b$ in the (010)-oriented samples and an expansion of the out-of-plane interplanar distance $c \sin \beta$ for the (001)-oriented sample, where $\beta$ is the angle between the [100] and [001] directions. In a monoclinic lattice, the normal to the (001) plane is not parallel to the [001] direction. In contrast, because [010] is the unique axis of $\beta$-Ga$_2$O$_3$, it is strictly perpendicular to the (010) plane.

In both samples, the absolute value of the perpendicular strain — which is calculated as $\varepsilon_\perp = (d - d_0)/d_0$, where $d_0$ and $d$ are the interplanar distances before and after the implantation, respectively — increases steadily with the fluence, while the crystalline quality, quantified by the DW factor, decreases. This evolution is well captured by the MROX fits, which show a very good agreement with the measured patterns, including all the Pendellösungen fringes. Remarkably, at each fluence, the magnitudes of the maximum observed strains are similar for all orientations, while the DW seems to achieve lower values in the case of the (001)-oriented sample. Moreover, both profiles seem to closely follow the vacancy profiles obtained by Stopping and Ranges of Ions in Matter (SRIM) Monte Carlo simulations[26], both regarding the affected depth and the general location of the maximum/minimum of the profiles. As such, this agreement emphasizes the interplay between the defects introduced during the ion implantation and the strain fields that accumulate in the material.

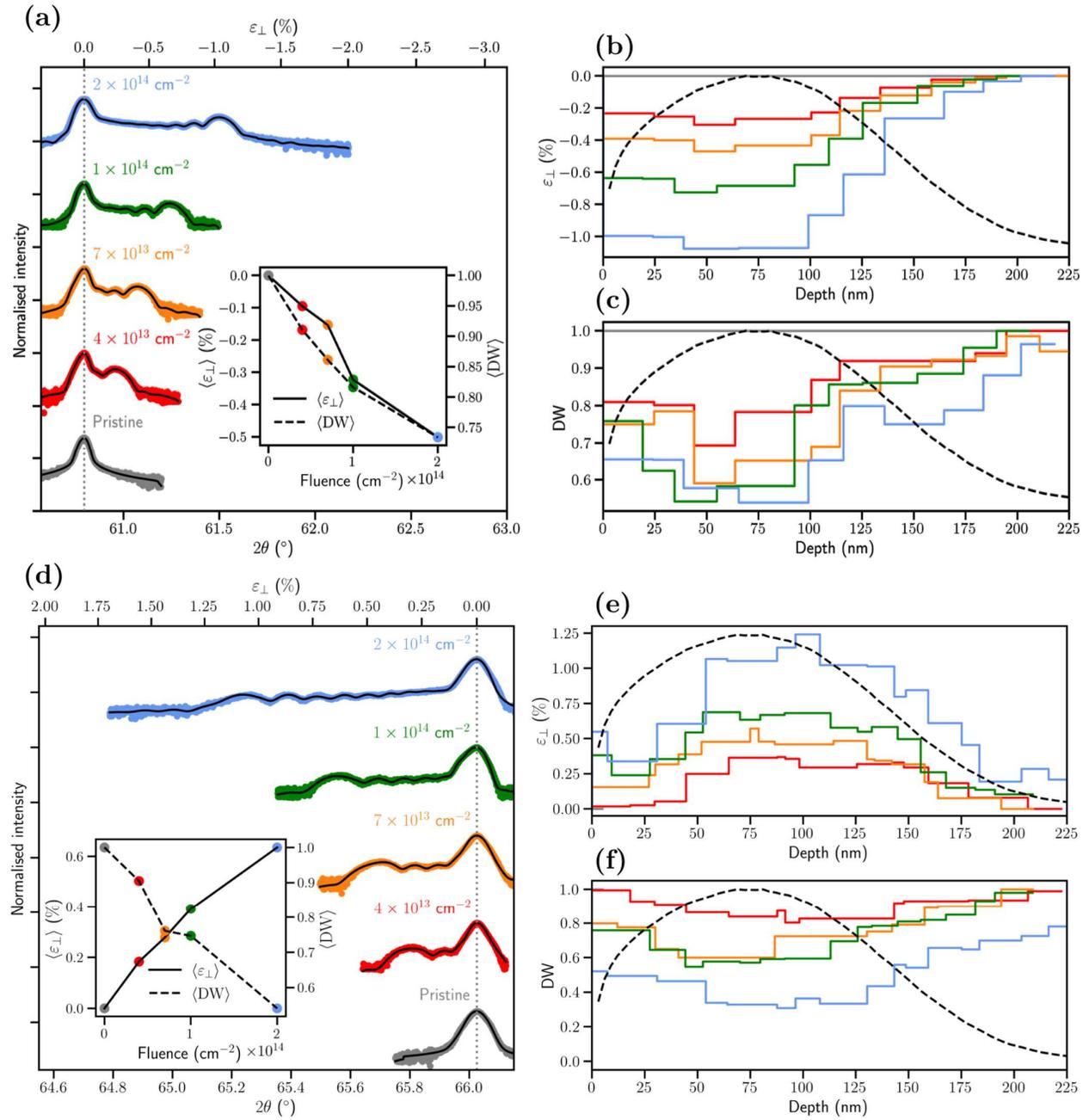

**Fig. 1** | X-ray diffraction analysis for the (010)- (a–c), about the 020 reflection, and (001)-oriented samples, (d–f), about the 004 reflection. Panels (a) and (d) show the experimental diffractograms (points) and the MROX simulations (black lines), in a logarithmic scale. The corresponding perpendicular strain, $\varepsilon_\perp$, profiles are shown in panels (b) and (e), while panels (c) and (f) show those of the Debye-Waller factors, DW. The dashed lines represent the vacancy profiles obtained from SRIM Monte Carlo simulations, plotted in an arbitrary scale. The insets show the average values of $\varepsilon_\perp$ and DW as a function of the implantation fluence.

It is also important to note that, while an accumulation of strain along the out-of-plane direction, as observed in Fig. 1, for both sample orientations, was previously reported by J. Matulewicz et al.[12], a full understanding requires the assessment of the in-plane strain as well. For this, reciprocal space maps (RSM) were acquired for selected symmetric and asymmetric reflections and did not reveal any strain along the in-plane directions.

Considering the relative intensity of the accessible reflections (both geometrically and kinematically), the selected reflections were 600, 710 and 80$\bar{1}$ for the (100)-oriented sample; 020, 110 and 022 for the (010)-oriented sample; and 004, 204 and 024 for the (001)-oriented sample. These RSM are shown in the supporting information, SI (Figs. S2–S4).

As the in-plane lattice parameters are found to remain fixed after implantation and strain accumulates only along the out-of-plane direction, the sample is considered to be subject to an in-plane stress state. In particular, this state results from the superposition of the stress field associated with the defects induced by ion implantation and the stresses imposed by the pristine part of the sample lying below the implanted region. In other words, as suggested previously[27,28], the stress introduced by the defects in the out-of-plane direction can be relaxed by a change in the out-of-plane lattice parameter, while the in-plane stress is fully compensated by the substrate to keep the in-plane lattice constants fixed. For a cubic crystal, A. Debelle and A. Declémy have previously shown that this substrate response term can be expressed as a function of the elastic constants of the material[28]; however, in a monoclinic system, the elasticity tensor has 13 independent components, and this calculation is not straightforward.

## 2.2. Stress vs strain anisotropic elasticity model

In order to further develop the results reported above, it is important to understand the physics behind the stress-strain relations in a monoclinic system under ion implantation. We start by considering that the stress-strain state of the implanted region results of a combination of two effects[29,30]: on the one hand, during the irradiation, a number of defects are introduced, displacing surrounding atoms and thus contributing to the strain/stress; on the other hand, the implanted region is also subject to interaction with the part of the sample that remains pristine (thus playing the role of a "substrate", with lattice constants $a_0$, $b_0$, $c_0$ and $\beta_0$, see Fig. S1 in the SI), which will also induce stress. Following the seminal work by J. Eshelby[31], we propose a model where the implanted region is considered on its own as a different material whose relaxed (without any external forces) lattice constants $a'$, $b'$, $c'$ and $\beta'$ differ from those of the pristine material due to the presence of the defects. Note that these also differ from the real situation (with lattice constants $a_i$, $b_i$, $c_i$ and $\beta_i$) due to the absence of the substrate.

The experimental RSM have shown that this interaction is such that the in-plane lattice constants of the pristine and implanted regions are the same, leading to a situation similar to pseudomorphic epitaxy of a thin film on a substrate. Fig. 2 shows a sketch of the situation for the (010) surface, as observed along the direction perpendicular to the (001) plane. With respect to the pristine situation (**1.**), the damaged layer (**2.**) is extended along the [100] direction and compressed along the [010] direction. However, since this layer is pseudomorphic to the substrate, as shown in situation **3.**, the latter must exert a compressive stress along [100] in order to keep the in-plane lattice parameter $a$ constant, which, in turn, leads to a slight expansion of $b$ via the Poisson effect. The associated (residual or latent) stresses can be determined from the pseudomorphic epitaxial relations deduced by M. Grundmann[32]. The situation is analogous along [001] and for the other surface orientations.

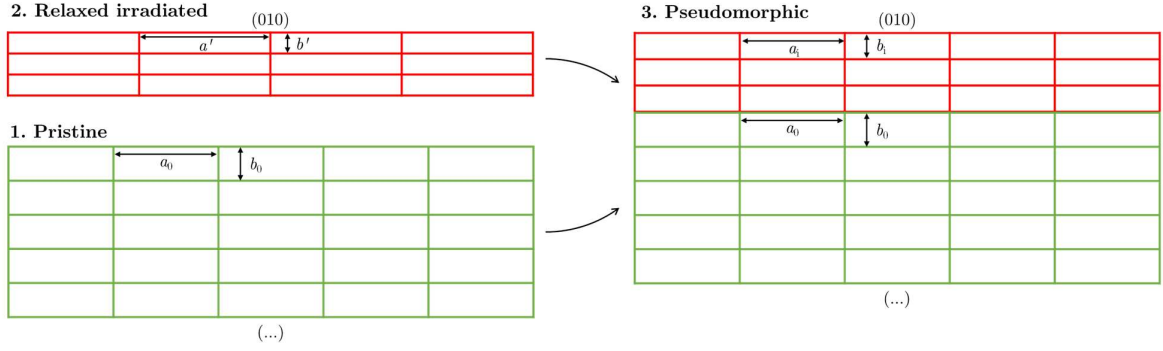

**Fig. 2** | Conceptual sketch of the evolution of lattice parameters with ion implantation. **1.** A pristine sample has an in-plane lattice constant $a_0$ and an out-of-plane constant $b_0$. **2.** The damaged layer, if free to relax, would have in- and out-of-plane lattice constants $a'$ and $b'$, respectively, with $a' > a_0$ and $b' < b_0$. **3.** In reality, the damaged layer is pseudomorphic to the pristine substrate, leading to $a_i = a_0$, while the out-of-plane constant increases slightly to $b_i > b'$ via the Poisson effect.

The stress-strain relation can be expressed as $\sigma_i = C_{ij}\varepsilon_j$, where $\boldsymbol{\sigma}$ and $\boldsymbol{\varepsilon}$ are the stress and strain tensors (in Voigt notation), respectively, and $\underline{C}$ is the stiffness matrix. Note that $\boldsymbol{\sigma}$ here represents the internal stresses on the film. For the monoclinic system, $\underline{C}$ can be written as[33]:

$$\underline{C} = \begin{pmatrix} C_{11} & C_{12} & C_{13} & 0 & C_{15} & 0 \\ & C_{22} & C_{23} & 0 & C_{25} & 0 \\ & & C_{33} & 0 & C_{35} & 0 \\ & & & C_{44} & 0 & C_{46} \\ & \text{sym} & & & C_{55} & 0 \\ & & & & & C_{66} \end{pmatrix}, \tag{1}$$

where "sym" denotes that the matrix is symmetric ($C_{ij} = C_{ji}$). In this work, we follow the numerical values determined experimentally by W. Miller et al.[33] for $\beta$-Ga$_2$O$_3$, after applying the appropriate rotations in order to express the tensor in the correct reference frame[32,34].

Following the notation in Fig. 2, we propose that the (macroscopic) strain can be written as:

$$\varepsilon = \varepsilon^* + e + \varepsilon^* e \simeq \varepsilon^* + e, \tag{2}$$

where $\varepsilon = (d_i - d_0)/d_0$ is the total strain, $\varepsilon^* = (d' - d_0)/d_0$ is known as an eigenstrain and $e = (d_i - d')/d'$ is the elastic strain, while $d$ plays the role of any lattice constant. The approximation is correct to the 1st order in the strain. Eigenstrains correspond to the strains induced by defects that would develop in the absence of external constraints and are used as stress-free reference strains; on the other hand, the elastic component arises from the stresses $-\sigma_j$ that are introduced in the implanted layer by the pristine region underneath to counteract the internal stresses $+\sigma_j$, and can be written in a Voigt-contracted form as $e_i = -C_{ij}^{-1}\sigma_j$[33]. The in-plane eigenstrains are exactly those that must be cancelled by the in-plane stresses introduced by the substrate to enforce a pseudomorphic condition.

To probe this model, classical MD simulations of the ion implantation into the different surface orientations were performed using 10 keV Ga atoms as projectiles, leading to a comparable number of dpa with respect to 250 keV Cr ions[15] at each fluence. The selected statistical ensemble for the region where the collision cascades developed was NVE (i.e., fixed number of particles, N, volume, V, and energy, E), with an NVT (i.e., fixed N, V and temperature, T) region underneath. This combination of fixed-volume ensembles with the employed boundary conditions — periodic in-plane and fixed in the out-of-plane direction — allows the role of the substrate to be emulated, including the heat dissipation aspect. Additional details regarding the MD simulation conditions can be found in the end matter section and in reference [15] for the (100) case. In order to probe the fluence evolution, 50 projectiles were sequentially simulated, leading to overlapping cascades.

After each cascade, the infinitesimal strain (with respect to the initial configuration), as given by Eq. (3), and the internal virial stresses, as given by Eq. (4) (see end matter), were computed for each atom and averaged over the damaged region. Note that the stresses calculated here are the internal stresses, i.e., they are opposite to those introduced in-plane by the substrate/container. Considering the stiffness tensor shown in Eq. (1), the elastic component of the strain $\underline{e}$ associated with the calculated stresses was also computed; from these, the eigenstrains were obtained as $\varepsilon^* = \varepsilon - e$. The component-resolved results are shown in Fig. 3 for ions impinging on the (100), (010) and (001) planes.

Considering the results for the infinitesimal strain, shown in Figs. 3 (a–c), it is quite clear that the dominant strain components for all surface orientations are those that lie out-of-plane (i.e., the solid lines in Fig. 3), namely the $xx$ component for (100), $yy$ for (010) and $zz$ for (001); this agrees with the experimental RSM shown in the SI, where no in-plane strains are observed. Moreover, their signs are also very consistent with the results in Fig. 1 and with the literature[12,15,25,35], i.e., the strain is negative (compressive) along the $b$-direction and positive (tensile) in the directions perpendicular to the (100) and (001) planes.

Regarding the stress accumulation, it is also quite clear that the dominant components are the ones in-plane (i.e., the dashed lines in Fig. 3), namely $yy$ and $zz$ for (100)-oriented cells, $xx$ and $zz$ for (010)-oriented cells and $xx$ and $yy$ for (001)-oriented cells; the out-of-plane and shearing components are thus deemed negligible in this context. Moreover, all the aforementioned stresses are tensile, with the exception of the $yy$ component, which is compressive. This is again in agreement with the measurements above, and reflects the natural tendency of the implanted layer to compress along the [010] direction. This fact has recently been exploited to produce microtubes that roll-up along this direction after the exfoliation of (100)-oriented bulk crystal[15].

For the (100) surface, Fig. 3 shows that the elastic strain along $b$ is positive, as the substrate compensates for the compressive internal stress, while the eigenstrain is negative; in the nomenclature of Fig. 2, $b_i = b_0 > b'$. On the other hand, the negative elastic strain along $c$ reflects the compressive stress introduced by the substrate to prevent the expansion if this lattice parameter, meaning that $c_i = c_0 < c'$. The combination of these in-plane stresses leads to the appearance of an additional elastic strain along the out-of-plane direction via the Poisson effect (even if that direction is stress-free), which is negative. As such, the final total strain corresponds to the sum of the defect-related eigenstrain and this elastic strain due to the reaction of the substrate. The interpretation for the other two orientations is analogous. However, it is curious to notice that the elastic strain along the direction perpendicular to the (001) plane in samples with that surface orientation is almost zero. In this case, the measured strain is, essentially, the eigenstrain introduced by the defects. Please refer to section S3 in the SI for a more thorough discussion of these results.

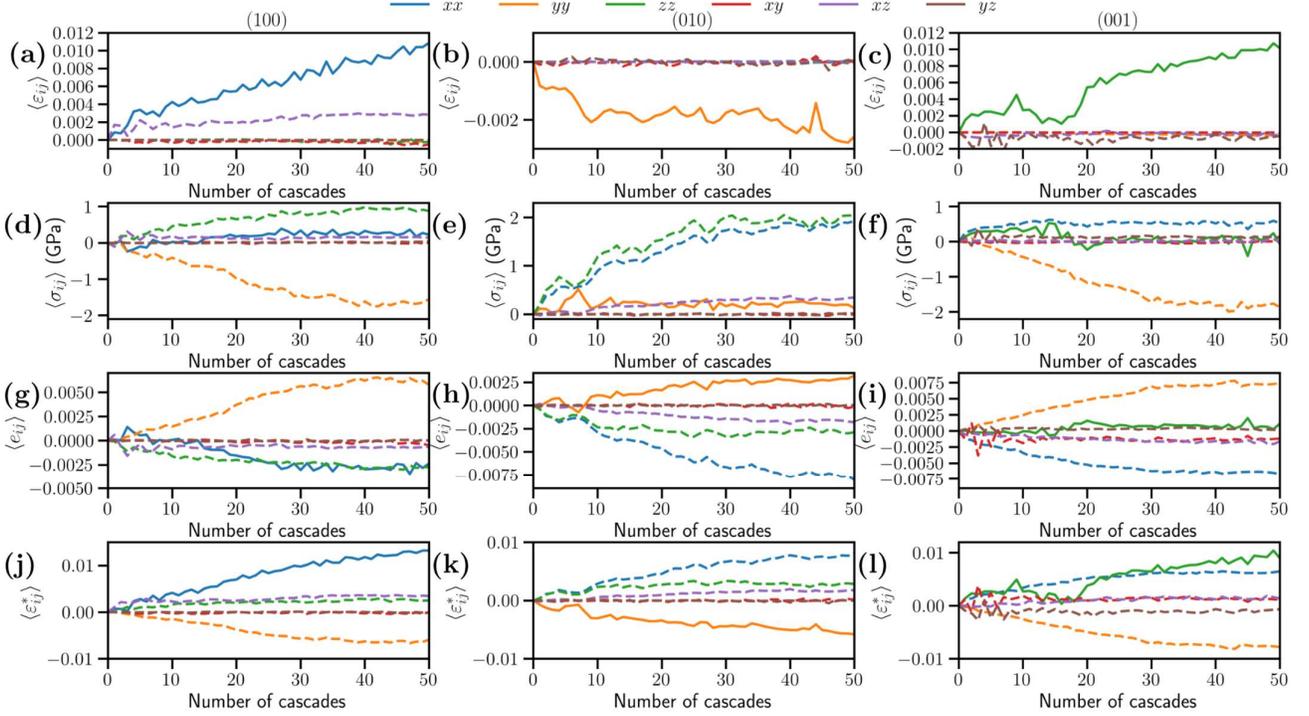

**Fig. 3** | MD-calculated average components of the infinitesimal strain (a–c), virial stress (d–f), elastic strain (g–i) and eigenstrain (j–l) tensors, obtained as a function of the number of overlapping cascades. Panels (a), (d), (g) and (j) refer to (100)-oriented, (b), (e), (h) and (k) to (010)-oriented and (c), (f), (i) and (l) to (001)-oriented surfaces. The solid lines refer to out-of-plane components, while the dashed ones refer to the in-plane components. The used reference systems are such that $x \perp (100)$, $y \parallel [010]$ and $z \parallel [001]$ for the (100) and (010) surface orientations and $x \parallel [100]$, $y \parallel [010]$ and $z \perp (001)$ for the (001) surface orientation.

### 2.3. Higher-dpa implantation: the loss of anisotropy

To assess the anisotropy at higher dpa levels above threshold for the $\beta$-to-$\gamma$ phase transition[36], samples with (100), (010) and (001) surface orientations were implanted with 300 keV $Tb^{2+}$ ions to a fluence of $1.0 \times 10^{16}$ $cm^{-2}$, corresponding to a dpa of about 10 at the depth of maximum nuclear energy deposition (above the ~0.78 dpa threshold for the phase transformation). So as to further clarify the crystalline relation between the $\beta$- and the induced $\gamma$-phases, pole figures were acquired for both (010)- and (001)-oriented samples, before and after the ion implantation, for the $2\vartheta$ value expected for the 311 reflection of the $\gamma$-phase ($2\vartheta = 35.8°$), which was chosen because it is the most intense one, thus enhancing the measurement statistics.

The results are displayed in Fig. 4. In the case of the as-grown single crystals, shown in Figs. 4 (a) and 4 (e), the pole figures consist of isolated and sharp peaks. As indicated by the white arrows in Fig. 4, these peaks are compatible with the 111, 311, 310 and 401 (and equivalent) reflections of $\beta$-$Ga_2O_3$, confirming the single-crystalline nature of the samples. After the implantation, for both surface orientations, a number of broader and less intense structures were observed in addition to the aforementioned peaks. These new structures were found to be fully compatible with the formation of a single-crystalline $\gamma$-phase region with a well-defined orientation. In fact, from this measurement, it was possible to infer the crystalline relation between the $\beta$- and $\gamma$-phases: $(0\bar{1}0)_\beta \parallel (110)_\gamma$ and $[102]_\beta \parallel [1\bar{1}2]_\gamma$, which was satisfied for both the (010)- and (001)-oriented samples, in agreement with previous works[37]. This justifies the appearance of the 440 $\gamma$-phase reflection on the $2\vartheta$–$\omega$ scans performed about the 020 reflection for the (010)-oriented sample, as the corresponding planes are parallel to the surface (see Fig. S5 (a) in the SI). On the other hand, it also justifies why no additional peaks were observed on the $2\vartheta$–$\omega$ scans performed about the 002 reflection (see Fig. S5 (b) in the SI) for the (001)-oriented sample, as the low-index $\gamma$-phase plane closest to parallel to the (001) planes (~1.2° tilt) of the $\beta$-phase is $(5\bar{5}2)$; however, this reflection is forbidden, rendering the $\gamma$-phase essentially undetectable by means of conventional symmetric $2\theta - \omega$ scans[12].

In short, these results show a clear preferential arrangement between the two phases, which is independent of the orientation of the surface. For the (100)-oriented samples, the formation of microtubes was observed, as expected for such a high dpa[15], which hinders the XRD analysis of the implantation damage, since the implanted region is removed. However, if the relation above also holds for this orientation, the (001) plane of the $\gamma$-phase should be almost parallel to the surface (~1.2° tilt). For the commonly-used $(\bar{2}01)$ orientation, the surface should be parallel to the $\{1\bar{1}\bar{1}\}$ family of planes, which is also in agreement with recent reports[38].

This $\beta$-to-$\gamma$ phase transformation has been previously observed in MD simulations, where its identification was made by comparing the 2nd coordination shell of the Ga–Ga radial distribution functions[11], which is a real-space method. In this context, a higher-fluence MD simulation was performed in a smaller cell in order to replicate the formation of the $\gamma$-phase, in agreement with a previous work[11] (see the simulation details in the end matter); remarkably, by employing Eqs. (4) and (5), it was possible to reproduce the experimental pole figures from these simulation cells. Additionally, Figs. S6 and S7 in the SI show an example of how this method can be used to simulate electron diffraction patterns that agree with previous experimental results[11,39], aiding in phase identification allowing the design of well-informed strategies for future experiments. By computing diffraction patterns directly from atomistic (or even *ab initio*) simulations, the method employed here also circumvents the issues related with the fact that the link between diffraction data and the defects remains largely model-dependent (e.g. the DW and defect profiles shown in section 2.1.).

For the pristine cells, the pole figures shown in Figs. 5 (c) and 5 (g) have well-defined and sharp peaks that are compatible with the indicated reflections. These pole figures thus confirm the surface orientations and also the azimuthal positioning of the cells to agree with the experiments. Once the cells are implanted, several broad, but still well-defined peaks emerge, that are compatible with reflections associated with the $\gamma$-phase, as shown in Figs. 4 (d) and 4 (h). Note that the broadening of diffraction peaks may be associated both with the small volume of the implanted region or with the presence of defects, which introduce inhomogeneous strain. Since the simulation cells have constant volumes and are damaged approximately uniformly, the $\beta$-related peaks disappear, and the broadness of the $\gamma$-related peaks is likely due to defects. In the experimental case, only part of the sample is implanted, so the $\beta$-related peaks remain. In any case, one obtains an excellent agreement between the experimental and simulated pole figures.

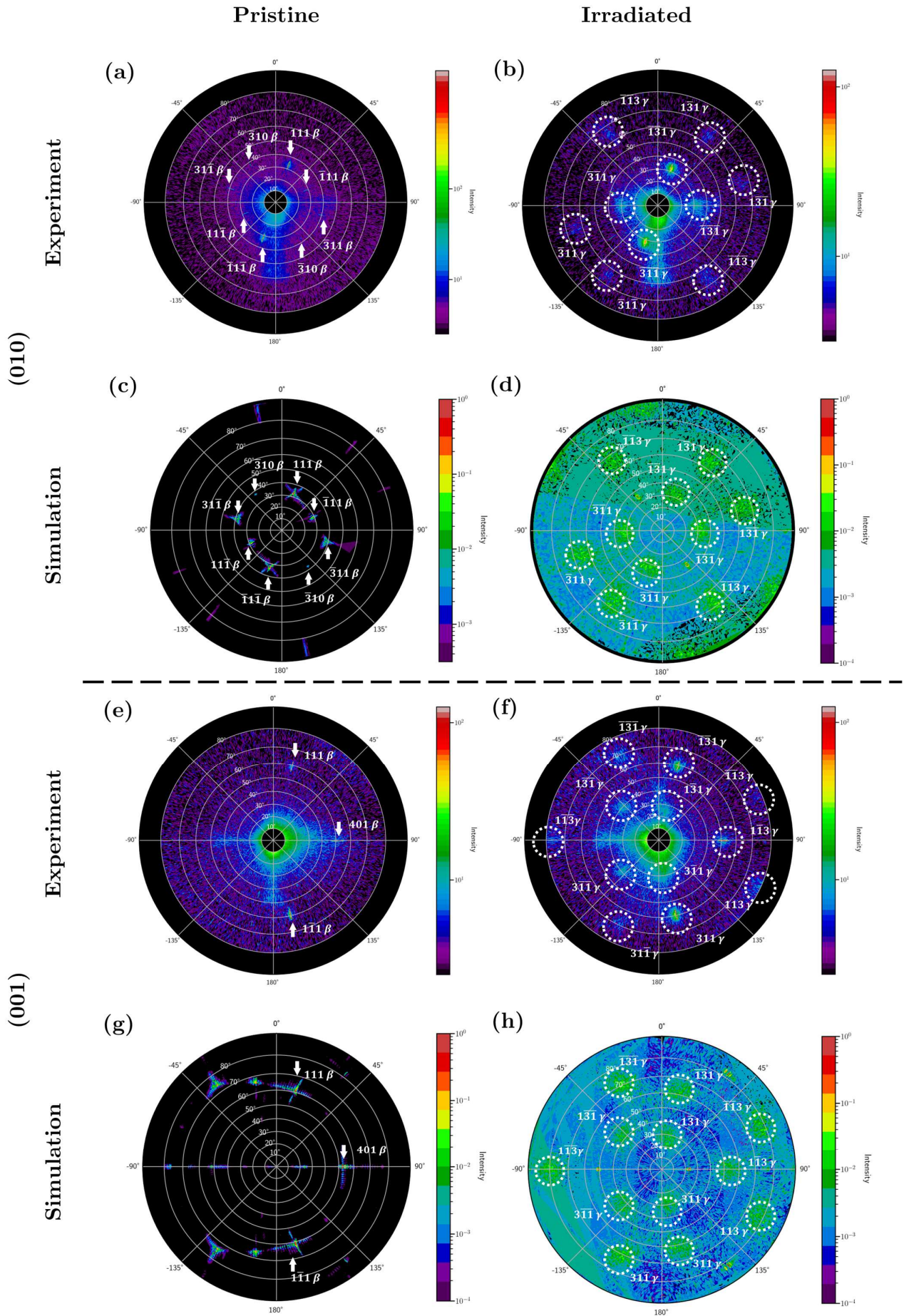

**Fig. 4** | Pole figure analysis of the β- to γ-phase transition under Tb-implantation. Panels (a–d) refer to the (010) surface orientation, while (e–h) refer to (001). Panels (a)/(e) show the experimental pole figures obtained before the implantation, while (b)/(f) show the same after the implantation. Panels (c)/(g) show the simulated pole figures obtained before the implantation, while (d)/(h) show the same after the implantation.

In short, the transformation from the monoclinic to the defective spinel (cubic) phase upon ion implantation, which involves an increase of the lattice symmetry, does not seem to have any preference regarding the surface orientation and always occurs in such a way that the relation $(0\bar{1}0)_\beta \parallel (110)_\gamma$ and $[102]_\beta \parallel [1\bar{1}2]_\gamma$ is fulfilled.

## 3. Conclusions

In this work, we assessed the structural and elastic properties of β-$Ga_2O_3$ under ion implantation performed into samples with different surface orientations, namely (100), (010) and (001). Strain and crystalline quality profiles were successfully extracted from XRD patterns via simulations employing the dynamic theory of X-ray diffraction. We show a clear difference in the strain accumulation in each of these surfaces, as the strain is found to be compressive in nature along the [010] and tensile along the direction perpendicular to (100) and (001).

In order to understand the atomistic origin of these stresses/strains introduced by ion implantation in β-$Ga_2O_3$, MD simulations were performed. The infinitesimal strain and virial stress were assessed as a function of the fluence, considering implantations performed into cells with the three main surface orientations: (100), (010) and (001). The results paint a consistent picture across the different surfaces and are in excellent agreement with experiment, revealing an accumulation of strain in the out-of-plane direction, while the sample is subjected to in-plane residual stresses. Specifically, the competition between the stresses introduced by the defects and those applied by the pristine regions of the sample leads to a zero-strain condition in-plane, in a physical situation similar to pseudomorphic epitaxy. The substrate reaction enforces a zero in-plane strain condition, while the out-of-plane direction is free to relax, leading to strain; additionally, the strain along the latter direction is affected by the in-plane stresses via the Poisson effect. As such, the modeling results are consistent with the proposed physical model for the anisotropic elasticity.

Finally, by probing a fluence regime above the threshold for the β-to-γ phase transition, we showed that this is an orientation-independent transformation that occurs in a well-ordered manner. It promotes an enhancement of the symmetry (monoclinic to cubic) and the crystallographic relation between the two polymorphs is $(0\bar{1}0)_\beta \parallel (110)_\gamma$ and $[102]_\beta \parallel [1\bar{1}2]_\gamma$. We also proposed to simulate the pole figures and other diffraction patterns from MD simulations cells, which allows a direct comparison between experimental diffraction patterns and MD simulations. Such forward simulation scheme is readily transferable to other materials, processing techniques and topics, including ultrafast non-equilibrium physics or in-situ phase-changes.

## 4. End matter

In this work, unintentionally doped commercial $\beta$-Ga$_2$O$_3$ single-crystals from Novel Crystal Technology were used. These crystals were grown by the edge-defined film-fed growth (EFG) method, cut along the (100), (010) or (001) planes, yielding 5 mm × 5 mm samples with a thickness of ~500 μm.

The ion implantations with 250 keV Cr$^+$ were performed at room temperature at the 500 kV implanter of the Ion Beam Centre of the Helmholtz Zentrum Dresden-Rossendorf (IBC-HZDR), with fluences between $6.0 \times 10^{12}$ and $2.0 \times 10^{14}$ cm$^{-2}$, with a tilt angle of 7° to avoid channeling. The 300 keV Tb$^{2+}$ implantations, performed at the 210 kV high flux ion implanter of the Laboratory of Accelerators of Instituto Superior Técnico (IST), Universidade de Lisboa[40] with a fluence of $1.0 \times 10^{16}$ cm$^{-2}$, also with a tilt angle of 7°.

The HRXRD measurements were performed at the Bruker D8 Discover diffractometer of the Laboratory of Accelerators of IST, for both the symmetric $2\theta - \omega$ scans and the (a)symmetric reciprocal space maps. For HRXRD, the primary beam optics consists of a Göbel mirror, a 0.2 mm collimation slit, and a 2-bounce (220)-Ge monochromator, to select the copper (Cu) K$\alpha_1$ X-ray line (wavelength of 1.5406 Å). The secondary beam path entails a 0.1 mm slit and a scintillation detector. For the pole figures, a low-resolution configuration was employed, without the monochromator, in order to enhance the intensity of the signal from the implanted region. Moreover, the 0.1 mm slit was substituted by long Soller slits, which reduce the angular divergence of the diffracted beam, leading to a higher signal-to-noise ratio.

The Stopping and Range of Ions in Matter (SRIM) Monte Carlo simulations[26] were performed in the full damage cascades calculation mode, with displacement energies of 28 eV for Ga and 14 eV for O atoms[41] and a density of 5.88 g/cm$^3$ for $\beta$-Ga$_2$O$_3$[42].

The Classical MD simulations were performed using the Large-scale Atomic/Molecular Massively Parallel Simulator (LAMMPS) code[43], employing a tabulated Gaussian approximation interatomic potential (tabGAP)[44] for Ga$_2$O$_3$[17]. The simulation cells for the (100) and (001) orientations consisted of 320000 atoms, while the cell for the (010) surface orientation had 480000 atoms, in order to account for the enhanced channeling along the [010] channel. Periodic boundary conditions were employed along in-plane directions, while the out-of-plane directions were subjected to fixed boundary conditions. Following our previous work[15], all cells had, along the out-of-plane direction, a 10 Å thick layer of fixed atoms, followed by a 20 Å thick NVT layer (canonical ensemble), which was modelled with a Nosé-Hoover thermostat at 300 K[45], while the remaining cell was NVE (microcanonical ensemble). The projectiles consisted of Ga atoms that were initialized at random positions in the plane 20 Å above the surface with a kinetic energy of 10 keV. A tilt angle of 7° was employed to suppress the channeling effect. The simulation was then run for 20 ps using an adaptive time step[46], i.e., a time step that is chosen so that the energy and position of the atoms do not vary more than a given amount, based on the current values for the velocities and forces. Afterwards, the simulation is run for an additional 20 ps using the NVT ensemble at 300 K on the implantation region, with a time step of 0.001 ps. The electronic stopping of the ions was modelled as a friction term. The structure visualization was done using the Open Visualization Tool (OVITO)[47].

The calculation of the atomic infinitesimal strain tensor was implemented using Ovito, according to the displacement between the reference and final configurations[47].

$$\varepsilon_{ij} = \frac{1}{2}\left(\frac{\partial u_i}{\partial X_j} + \frac{\partial u_j}{\partial X_i}\right), \qquad (3)$$

where $u_i$ and $u_j$ denote the $i$th and $j$th components of the displacement vector and $X_i$ and $X_j$ denote the $i$th and $j$th components of the vector position in the coordinates of the reference (pristine) configuration. The average strain was calculated over the region approximately between 20 and 120 Å below the surface.

The virial stress tensor associated with each atom was calculated with LAMMPS, as follows:[48–50]

$$\sigma_{ij} = \frac{1}{V}\left[mv_i v_j + \frac{1}{2}\sum_n R_i(n)\, F_j(n)\right], \qquad (4)$$

where $m$ is the mass of the atom, $v_i$ and $v_j$ denote the $i$th and $j$th components of its velocity, $R_i(n)$ denotes the $i$th component of the (relative) position vector of the $n$th neighbour with respect to the considered atom and $F_j(n)$ denotes the $j$th component of the force exerted on the considered atom by its $n$th neighbour. The averaged stress was calculated over the region approximately between 20 and 120 Å below the surface, with a volume $V$, and time-averaged over 5 ps.

Note that the strain and stress components in this work are expressed with respect to an orthonormal basis $\{e_1, e_2, e_3\}$, where $e_2$ and $e_3$ are unit vectors along the $b$- and $c$-axes, respectively, while $e_1$ is a unit vector parallel to the reciprocal lattice basis vector $a^*$, i.e., it is perpendicular to the (100) plane, in the case of the (100)- and (010)-oriented cells; for the (001)-oriented cells, the base was such that $e_1$ and $e_2$ lie along the $a$- and $b$-axes, respectively, while $e_1$ is a unit vector parallel to the reciprocal lattice basis vector $c^*$, i.e. it is perpendicular to the (001) plane. Moreover, the calculated quantities are internal stresses, and the convention used is such that a positive stress is tensile, while a negative stress is compressive.

For higher dpa levels, overlapping cascade simulations were performed with a different simulation scheme employing periodic boundary conditions, as explained in our previous work[15]. These simulations used a $\beta$-$Ga_2O_3$ cell containing 81920 atoms and, in each iteration, a Ga or O atom is randomly selected as the primary knock-on atom (PKA). To ensure consistency, the entire cell is translated and wrapped at periodic boundaries, positioning the PKA at the center of the cell. The PKA is assigned a kinetic energy of 1.5 keV, with a uniformly random momentum direction. The collision cascade process is simulated within the microcanonical ensemble (NVE) using the same adaptive time-step algorithm. Electron-stopping frictional forces are applied to atoms with kinetic energies above 10 eV during the first 5000 steps. Afterwards, the simulations proceed in a quasi-canonical ensemble, where a Langevin thermostat is applied to atoms within 7.5 Å of the simulation box boundaries (which are redefined at each iteration) for 10 ps at 300 K. To release the stress in the defected cell, an additional relaxation step is carried out in the isothermal-isobaric ensemble (fixed number of particles, pressure, $P$, and temperature — NPT) at 0 bar and 300 K for 2 ps.

The reciprocal space of the cells was simulated from their atomic configurations under the Born and the kinematic approximations of X-ray diffraction; in this context, the scattering intensity of a distribution of $N$ scattering centres corresponding to the reciprocal space vector $\mathbf{Q} = (Q_x, Q_y, Q_z)$ is given by[19,51]:

$$I(\mathbf{Q}) \propto \left| \sum_{n=1}^{N} f_n(|\mathbf{Q}|) \exp(i \mathbf{Q} \cdot \mathbf{r}_n) \right|^2, \tag{5}$$

where $\mathbf{r}_n$ is the position of the $n$th atom in the simulation cell and $f_n(|\mathbf{Q}|)$ is the atomic scattering factor for that atom. In the present work, this factor was computed from the Cromer-Mann parametrization with the tabulated coefficients for $Ga^{3+}$ cations[52] and $O^{2-}$ anions[53]. The $\mathbf{Q}$ vector is then swept and the quadruples $(Q_x, Q_y, Q_z, I(\mathbf{Q}))$ are tabulated in the desired reference frame. By choosing the suitable frame and performing the appropriate cuts, the components of $\mathbf{Q}$ that are parallel ($Q_\parallel$) or perpendicular ($Q_\perp$) to a given surface can be calculated, thus allowing the comparison with results obtained from experimental RSM, as shown before[15]. It is interesting to notice that the $\mathbf{Q}$ domain will always be finite, since the maximum value of $|\mathbf{Q}|$, corresponding to the limit of the Ewald sphere, is $4\pi/\lambda$, where $\lambda$ is the wavelength of the probing radiation (e.g. X-ray or electron)[54], thus contributing to the computational efficiency. These simulations can also be used to assess phases, lattice constants and the angles between different planes. By simulating a larger-area RSM, this approach also allows the simulation of zone axis patterns, even if they are typically geometrically forbidden in an experiment (e.g., the zone axis perpendicular to the surface plane). The evolution with the number of overlapping cascades from the monoclinic to the defective spinel phases is shown in the SI (see Figs. S6 and S7), revealing a very good agreement with the experimental diffraction patterns obtained in previous works[11,39].

A third approach based on these calculations is producing stereographic projections of the reciprocal space that can be directly compared with pole figures, as illustrated in this paper. We start by converting the previously-simulated reciprocal space into spherical coordinates $(|\mathbf{Q}|, \theta, \varphi, I(\mathbf{Q}))$, where $\theta$ and $\varphi$ are the polar and azimuthal angle, respectively. Note that this means that the projection is made with respect to the surface plane perpendicular to the $z$-axis; if another plane is desired, the simulation cell must be rotated in order to meet that condition. The rotation angle and axes may be conveniently calculated, e.g., using our algebraic and numerical Mathematica notebook for crystallographic computations[55]. Since, experimentally, a pole figure is performed in a small range about a given value of $2\vartheta$ (corresponding to a range of interplanar distances $d$), this corresponds to selecting a range of radial coordinates in reciprocal space, $[Q_{\min}, Q_{\max}]$, as $|\mathbf{Q}| \propto d^{-1}$. This will actually correspond to a spherical shell over which the intensity should be averaged, resulting in ordered triplets $(\theta, \varphi, \langle I_{\text{pole figure}}(\theta, \varphi) \rangle)$, where:

$$I_{\text{pole figure}}(\theta, \varphi) = \langle I(\theta, \varphi) \rangle = \frac{1}{Q_{\max} - Q_{\min}} \int_{Q_{\min}}^{Q_{\max}} I(|\mathbf{Q}|, \theta, \varphi, I) \, d|\mathbf{Q}|. \tag{6}$$

Finally, the stereographic projection of the northern hemisphere (considering the convention where the projection is done onto the equatorial plane, with respect to the south pole) is simply given by $(R, \Theta) =$ $(\tan(\theta/2), \varphi)$, where $R$ is the radial coordinate and $\theta$ is the angular one. Alternatively, for a powder pattern, one can perform a spherical average and obtain the diffraction pattern as a function of the radial coordinate $|\mathbf{Q}|$, which can be directly converted to $2\vartheta$ via Bragg's law, $2\theta = 2\arcsin\left(\frac{\lambda|\mathbf{Q}|}{4\pi}\right)$; this is similar to the approach taken by the Debyer software[56], i.e.:

$$I_{\text{powder}}(Q) = \langle I(|\mathbf{Q}|) \rangle = \frac{1}{4\pi} \int_0^{2\pi} \int_0^{\pi} I(|\mathbf{Q}|, \theta, \varphi) \, |\mathbf{Q}| \sin\theta \, d\theta \, d\varphi. \tag{7}$$

The code implementing these different aspects is freely available online on GitHub: https://github.com/DuarteME/RSM-MD

### Acknowledgements

The authors acknowledge the financial support from the Portuguese Foundation for Science and Technology (FCT) via the IonProGO project (2022.05329.PTDC, http://doi.org/10.54499/2022.05329.PTDC) and via the INESC MN Research Unit funding (UID/05367/2020) through Pluriannual BASE and PROGRAMATICO financing. The authors also acknowledge the high-performance computing resources provided by the Navigator platform of the Laboratory for Advanced Computing at University of Coimbra and the Deucalion supercomputer of the University of Minho, under the MDGaO FCT advanced computing projects (2023.09323.CPCA, https://doi.org/10.54499/2023.09323.CPCA.A1, 2023.11548.CPCA, https://doi.org/10.54499/2023.11548.CPCA.A1 and 2025.00028.CPCA.A2), as well as the Finnish IT Center for Science (CSC). D. M. Esteves (2022.09585.BD, https://doi.org/10.54499/2022.09585.BD) and J. Zanoni (2022.14486.BD, https://doi.org/10.54499/2022.14486.BD) thank FCT for their PhD grants. J. Zanoni, J. Rodrigues and T. Monteiro acknowledge financial support from the FCT via the project POEMS (101217446), which is supported by Chips Joint Undertaking, as well as the i3N projects UIDB/50025/2020 & UIDP/50025/2020 & LA/P/0037/2020. J. Rodrigues also acknowledges FCT support via grant CEECINSTLA/00005/2022. The Cr implantations were performed under proposal 26001 of the ReMade@ARI project (https://doi.org/10.3030/101058414), funded by the European Union as part of the Horizon Europe call HORIZON-INFRA-2021-SERV-01 under grant agreement number 101058414 and co-funded by UK Research and Innovation (UKRI) under the UK government's Horizon Europe funding guarantee (grant number 10039728) and by the Swiss State Secretariat for Education, Research and Innovation (SERI) under contract number 22.00187. Neither the European Union nor any of the granting authorities can be held responsible for those opinions.

# Supporting Information

# Understanding the anisotropic response of $\beta$-Ga$_2$O$_3$ to ion implantation

D. M. Esteves[1,2,*], R. He[3], S. Magalhães[2,4], M. C. Sequeira[5], Â. R. G. Costa[6], J. Zanoni[7], J. Rodrigues[7], T. Monteiro[7], F. Djurabekova[3], K. Lorenz[1,2,4], M. Peres[1,2,4]

[1] INESC Microsystems and Nanotechnology, Rua Alves Redol 9, Lisboa 1000-029, Portugal

[2] Institute for Plasmas and Nuclear Fusion, Instituto Superior Técnico, University of Lisbon, Av. Rovisco Pais 1, Lisboa 1049-001, Portugal

[3] Department of Physics, University of Helsinki, P.O. Box 43, FI-00014, Helsinki, Finland

[4] Department of Nuclear Science and Engineering, Instituto Superior Técnico, University of Lisbon, Estrada Nacional 10, km 139.7, Bobadela 2695-066, Portugal

[5] Helmholtz-Zentrum Dresden-Rossendorf, Bautzner Landstraße 400, 01328 Dresden, Germany

[6] Centre for Nuclear Sciences and Technologies, Instituto Superior Técnico, University of Lisbon, Estrada Nacional 10, km 139.7, Bobadela 2695-066, Portugal

[7] i3N, Department of Physics, University of Aveiro, Campus Universitário de Santiago, Aveiro 3810-193, Portugal

* Corresponding author: [duarte.esteves@tecnico.ulisboa.pt](duarte.esteves@tecnico.ulisboa.pt)

**S1. The $\beta$-Ga$_2$O$_3$ structure**

In order to help understand the discussion in the main text, Fig. S1 shows a model of the $\beta$-Ga$_2$O$_3$ conventional cell with the main directions, namely [100], [010] and [001], in perspective. Note that the [010] direction is perpendicular to both [100] and [001], while the angle between the latter two is $\beta > 90°$. Note also the directions perpendicular to the (100) and (001) planes, which are tilted by $90° - \beta$ with respect to the [100] and [001] directions, respectively. The tensors in the main text are shown with respect to an orthonormal basis $\{e_1, e_2, e_3\}$ which is different for different orientations. Namely, for the the (100)- and (010)-oriented samples, $e_1 \perp (100)$, $e_2 \parallel [010]$ and $e_3 \parallel [001]$, while for the (001)-oriented samples $e_1 \parallel [100]$, $e_2 \parallel [010]$ and $e_3 \parallel\perp (001)$.

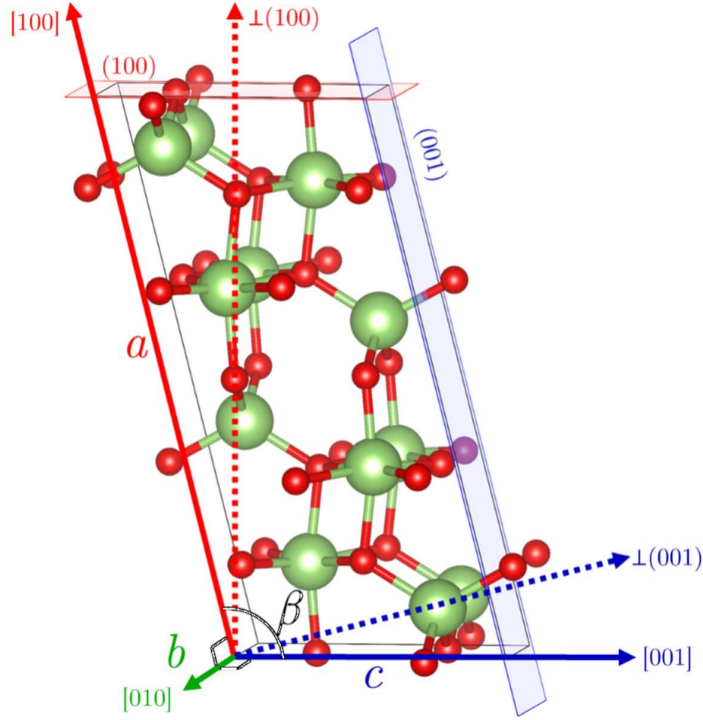

**Fig. S1** | Conventional unit cell for $\beta$-Ga$_2$O$_3$, where the Ga atoms are shown in green and the O atoms are shown in red. The red and blue planes correspond, respectively, to the (100) and (001) planes, and the dashed vectors of the same colour show the directions perpendicular to the respective plane. The figure was obtained using the Vesta software [1].

## S2. Experimental reciprocal space maps

Reciprocal space maps (RSM) of the (100)-, (010)- and (001)-oriented samples were measured about selected reflections, both in symmetric and asymmetric geometry, in order to probe the strain along different out-of-plane and in-plane directions. Considering the relative intensity of the accessible reflections (both geometrically and kinematically), the selected reflections were: $600$, $710$ and $80\bar{1}$ for the (100)-oriented sample; $020$, $110$ and $022$ for the (010)-oriented sample; and $004$, $204$ and $024$ for the (001)-oriented sample. These are shown in Figs. S2–S4 as a function of the fluence for a 250 keV Cr implantation, in order to assess the strain accumulation during implantation. In each figure, the perpendicular strain is calculated as $\varepsilon_\perp = (Q_{\perp 0} - Q_\perp)/Q_\perp$, where $Q_{\perp 0}$ and $Q_\perp$ are the perpendicular coordinates of the scattering vector $\boldsymbol{Q}$ before and after implantation, respectively. In each RSM, it is possible to observe a very intense peak which corresponds to the pristine regions of the sample. With increasing fluence, an additional structure appears, which reflects the strain accumulation in the sample. Just like in Fig. 1 of the main paper, there are clear differences among the orientations. For the (010)-oriented sample, there is a clear increase of $Q_\perp$, which indicates a compression of the out-of-plane lattice parameter, while for the (100)-oriented and (001)-oriented sample the value of $Q_\perp$ is reduced, suggesting an expansion of the out-of-plane lattice constant. This trend is consistent for all the probed reflections. In the asymmetric reflections, no shift of the peak is observed along the parallel component of $\boldsymbol{Q}$, $Q_\parallel$, suggesting that the in-plane lattice constants are fixed in all cases.

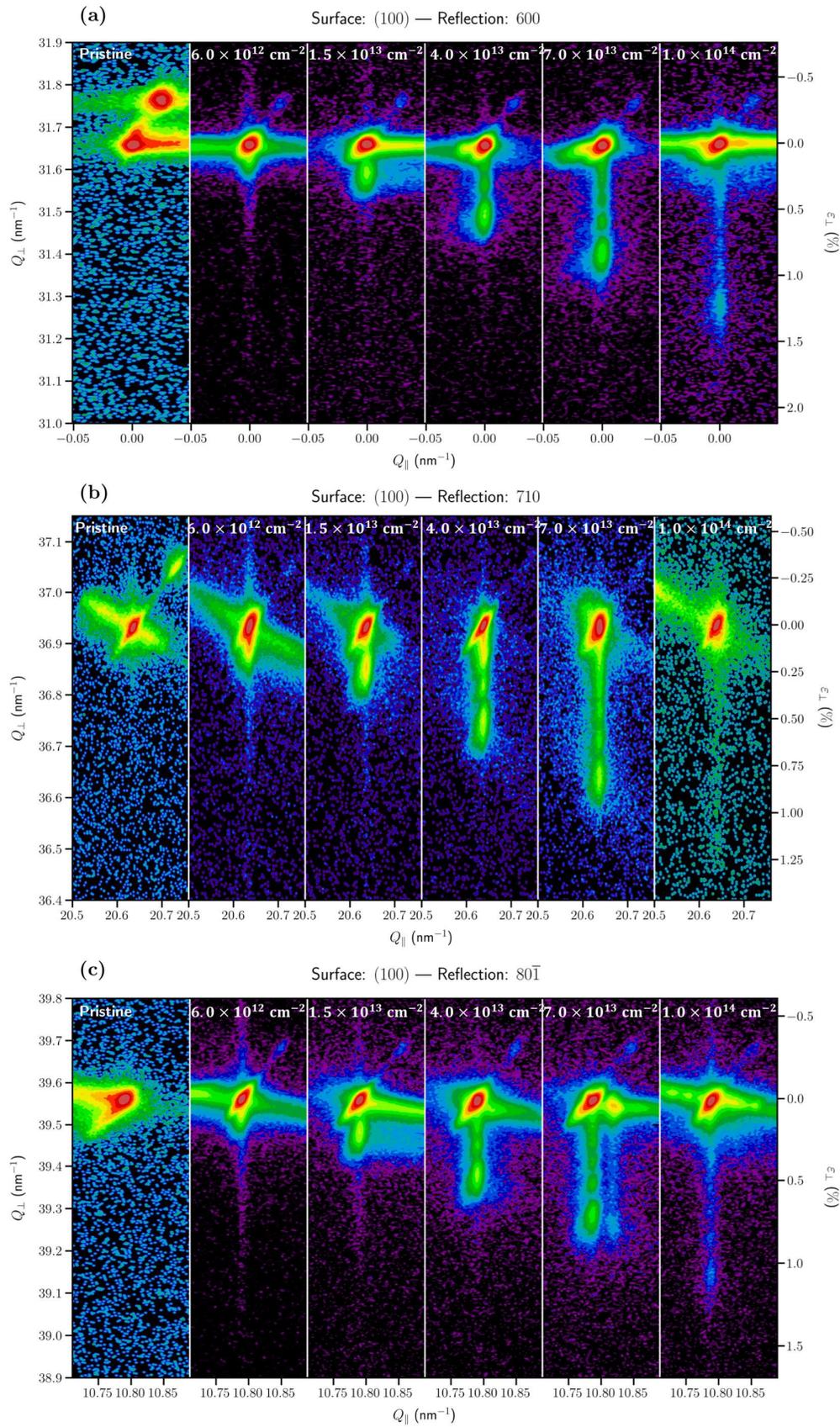

**Fig. S2** | Experimental RSM obtained for the (100)-oriented sample about the (a) **600**, (b) **710** and (c) **80$\bar{1}$** reflections, as a function of the implantation fluence. The additional feature on the upper-right hand side of the most intense peak corresponds to a contribution due to the diffraction of the Kα$_2$ line.

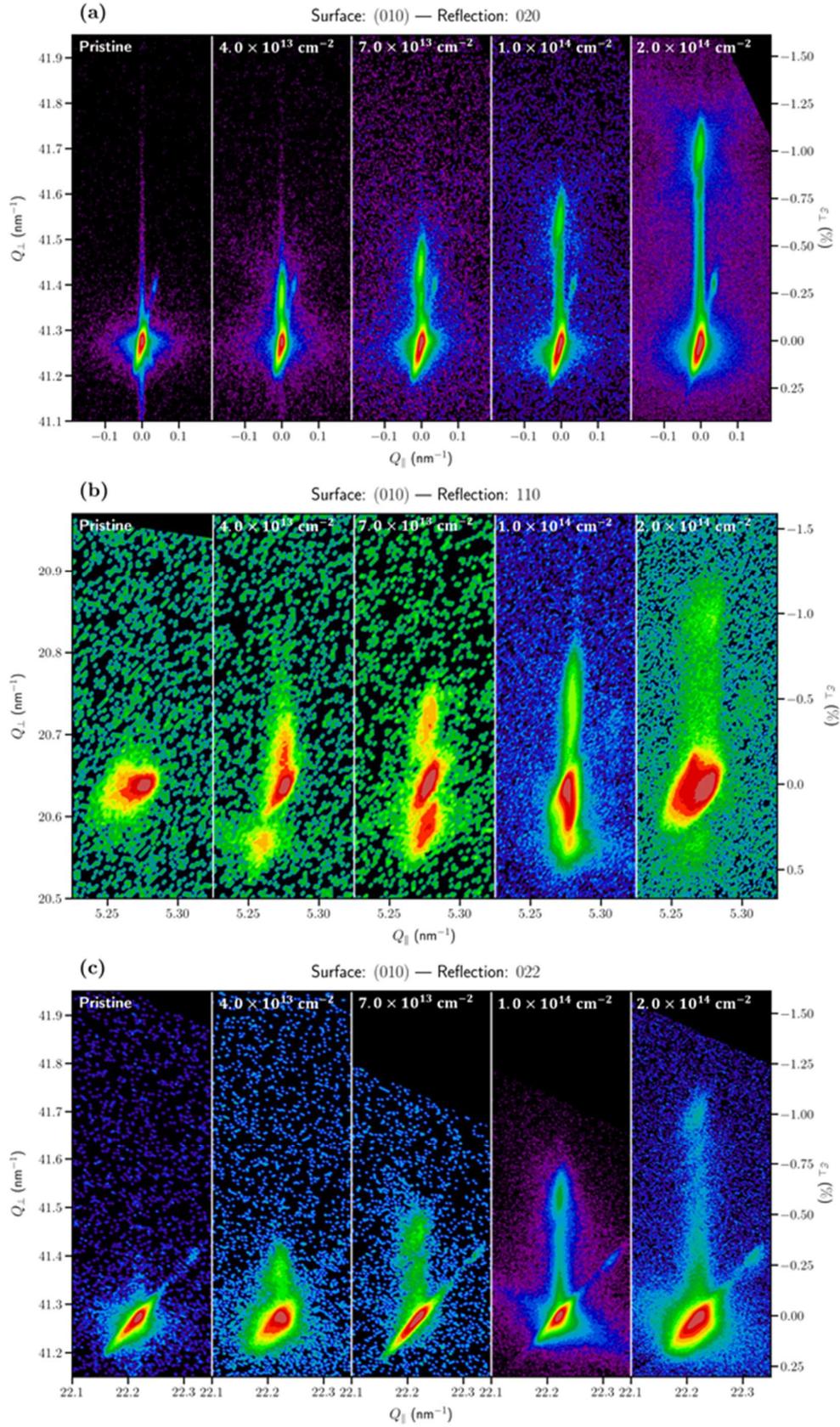

**Fig. S3** | Experimental RSM obtained for the (010)-oriented sample about the (a) **020**, (b) **110** and (c) **022** reflections, as a function of the implantation fluence. The additional feature on the upper-right hand side of the most intense peak corresponds to a contribution due to the diffraction of the Kα$_2$ line.

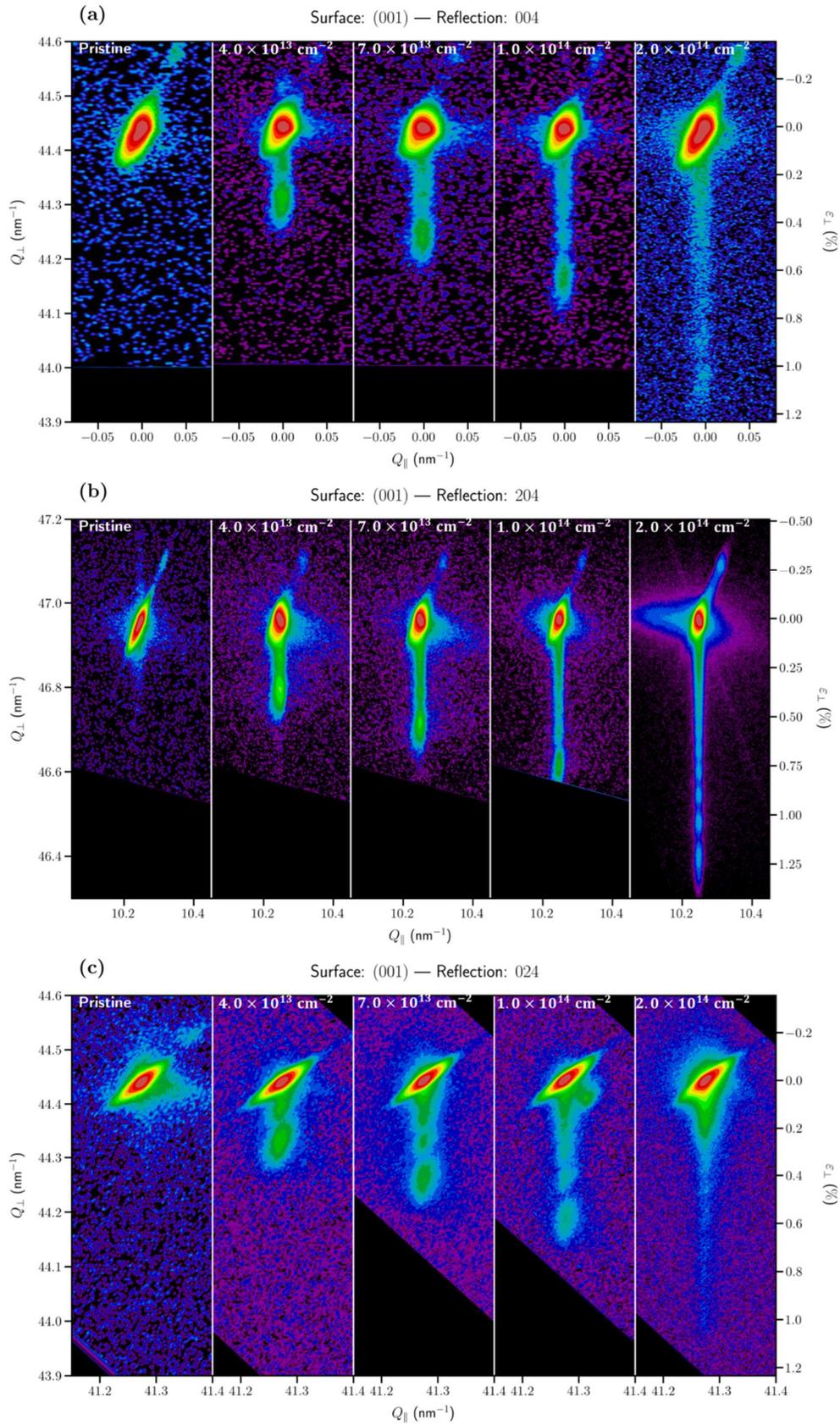

**Fig. S4** | Experimental RSM obtained for the (001)-oriented sample about the (a) **004**, (b) **204** and (c) **024** reflections, as a function of the implantation fluence. The additional feature on the upper-right hand side of the most intense peak corresponds to a contribution due to the diffraction of the K$\alpha_2$ line.

## S3. Stress/strain

In order to better understand the elasticity results shown in Fig. 3 of the main text, Tab. S1 shows the average values of each quantity between 40 and 50 overlapping cascades.

| Parameter/Surface | (100) | (010) | (001) |
|---|---|---|---|
| $\varepsilon_{xx}$ | $9.72\times10^{-3}$ | $1.48\times10^{-5}$ | $-1.43\times10^{-4}$ |
| $\varepsilon_{yy}$ | $-8.24\times10^{-5}$ | $-2.41\times10^{-3}$ | $-3.55\times10^{-4}$ |
| $\varepsilon_{zz}$ | $-1.53\times10^{-4}$ | $-1.88\times10^{-5}$ | $9.71\times10^{-3}$ |
| $\varepsilon_{xy}$ | $-3.89\times10^{-4}$ | $3.95\times10^{-6}$ | $0.00$ |
| $\varepsilon_{xz}$ | $2.85\times10^{-3}$ | $0.00$ | $-2.42\times10^{-4}$ |
| $\varepsilon_{yz}$ | $0.00$ | $-2.77\times10^{-5}$ | $-6.31\times10^{-4}$ |
| $\sigma_{xx}$ (GPa) | $2.66\times10^{-1}$ | $1.84$ | $5.13\times10^{-1}$ |
| $\sigma_{yy}$ (GPa) | $-1.66$ | $2.11\times10^{-1}$ | $-1.84$ |
| $\sigma_{zz}$ (GPa) | $9.28\times10^{-1}$ | $1.94$ | $3.41\times10^{-2}$ |
| $\sigma_{xy}$ (GPa) | $-2.15\times10^{-3}$ | $-6.48\times10^{-3}$ | $-7.53\times10^{-3}$ |
| $\sigma_{xz}$ (GPa) | $1.55\times10^{-1}$ | $3.19\times10^{-1}$ | $1.60\times10^{-2}$ |
| $\sigma_{yz}$ (GPa) | $1.82\times10^{-2}$ | $3.53\times10^{-3}$ | $1.27\times10^{-1}$ |
| $e_{xx}$ | $-2.77\times10^{-3}$ | $-7.56\times10^{-3}$ | $-6.45\times10^{-3}$ |
| $e_{yy}$ | $6.25\times10^{-3}$ | $2.76\times10^{-3}$ | $7.34\times10^{-3}$ |
| $e_{zz}$ | $-2.80\times10^{-3}$ | $-2.73\times10^{-3}$ | $7.94\times10^{-4}$ |
| $e_{xy}$ | $-1.88\times10^{-4}$ | $-4.00\times10^{-5}$ | $-1.24\times10^{-3}$ |
| $e_{xz}$ | $-6.88\times10^{-4}$ | $-1.66\times10^{-3}$ | $-1.69\times10^{-3}$ |
| $e_{yz}$ | $2.06\times10^{-5}$ | $3.25\times10^{-5}$ | $2.68\times10^{-4}$ |
| $\varepsilon^*_{xx}$ | $1.25\times10^{-2}$ | $7.57\times10^{-3}$ | $6.31\times10^{-3}$ |
| $\varepsilon^*_{yy}$ | $-6.33\times10^{-3}$ | $-5.17\times10^{-3}$ | $-7.70\times10^{-3}$ |
| $\varepsilon^*_{zz}$ | $2.65\times10^{-3}$ | $2.71\times10^{-3}$ | $8.92\times10^{-3}$ |
| $\varepsilon^*_{xy}$ | $-2.01\times10^{-4}$ | $4.39\times10^{-5}$ | $1.24\times10^{-3}$ |
| $\varepsilon^*_{xz}$ | $3.54\times10^{-3}$ | $1.66\times10^{-3}$ | $1.45\times10^{-3}$ |
| $\varepsilon^*_{yz}$ | $-2.06\times10^{-5}$ | $-6.02\times10^{-5}$ | $-8.99\times10^{-4}$ |

Tab. S1 | Average values of the stress, strain, elastic strain and eigenstrains shown in Fig. 3, between 40 and 50 overlapping cascades.

For each surface, we observe the following behaviour:

- **(100):** the dominant strain component is the out-of-plane ($xx$). The dominant internal stress components are in-plane ($yy$ and $zz$), with a compressive stress along $b$ and a tensile stress along $c$. These are compensated by the substrate, yielding a tensile elastic strain along $b$ and a compressive elastic strain along $c$. Since the signs are different along each direction, it is not a priori obvious how the Poisson effect will affect the out-of-plane direction; however, the elastic coupling between this direction and $b$ is larger than between this direction and $c$, leading to a small but negative elastic strain along $xx$; hence, the eigenstrain along this direction is slightly larger than the total strain.

- **(010):** the dominant strain component is the out-of-plane ($yy$). The dominant internal stress components are in-plane ($xx$ and $zz$), with compressive stresses along the direction perpendicular to (100) and $c$. These are compensated by the substrate, yielding compressive strain along those directions. The in-plane compression leads to a slight out-of-plane expansion via the Poisson effect, so the elastic strain along $b$ is positive. The eigenstrain is negative along $b$, and dominates over the elastic strain, leading to a compressive total strain. To a lesser degree, there is a shearing stress in the $x$–$z$ plane, which is compensated to keep the $\beta$ angle fixed.

- **(001):** this orientation is similar to (100). The dominant strain component is the out-of-plane ($zz$), while the dominant internal stress components are in-plane ($xx$ and $yy$), which is compressive stress along $b$ and tensile along $a$. The substrate reaction yields a tensile elastic strain along $b$ and a compressive elastic strain along $a$. In this case, the opposite signs lead to an almost zero elastic strain along the out-of-plane direction, and the total strain is essentially equal to the eigenstrain.

In general, the results consistently suggest that, under ion implantation, $Ga_2O_3$ tends to experience compressive internal stresses along $b$ and tensile stresses along $a$ and $c$, as well as in the directions perpendicular to the (100) and (001) planes. If relaxed freely, these stresses would lead to a compressive strain along the former direction and to tensile strain along the latter directions, which correspond to the eigenstrains. However, since the implanted layer is pseudomorphic to the substrate, additional stresses are induced in order to cancel these in-plane strains. In turn, these stresses also have a reflection on the out-of-plane lattice parameter via the Poisson effect. The degree to which each directions couple to one another through this effect depends on the elastic constants of the stiffness matrix.

### S4. Experimental $2\vartheta$–$\omega$ scans

After an implantation with 300 keV $Tb^{2+}$ ions to a fluence of $10^{16}$ cm$^{-2}$, corresponding to a dpa of about 10 at the depth of maximum nuclear energy deposition, the symmetric $2\theta - \omega$ scans were acquired for the (010)- and (001)-oriented samples, as shown in Fig. S5.

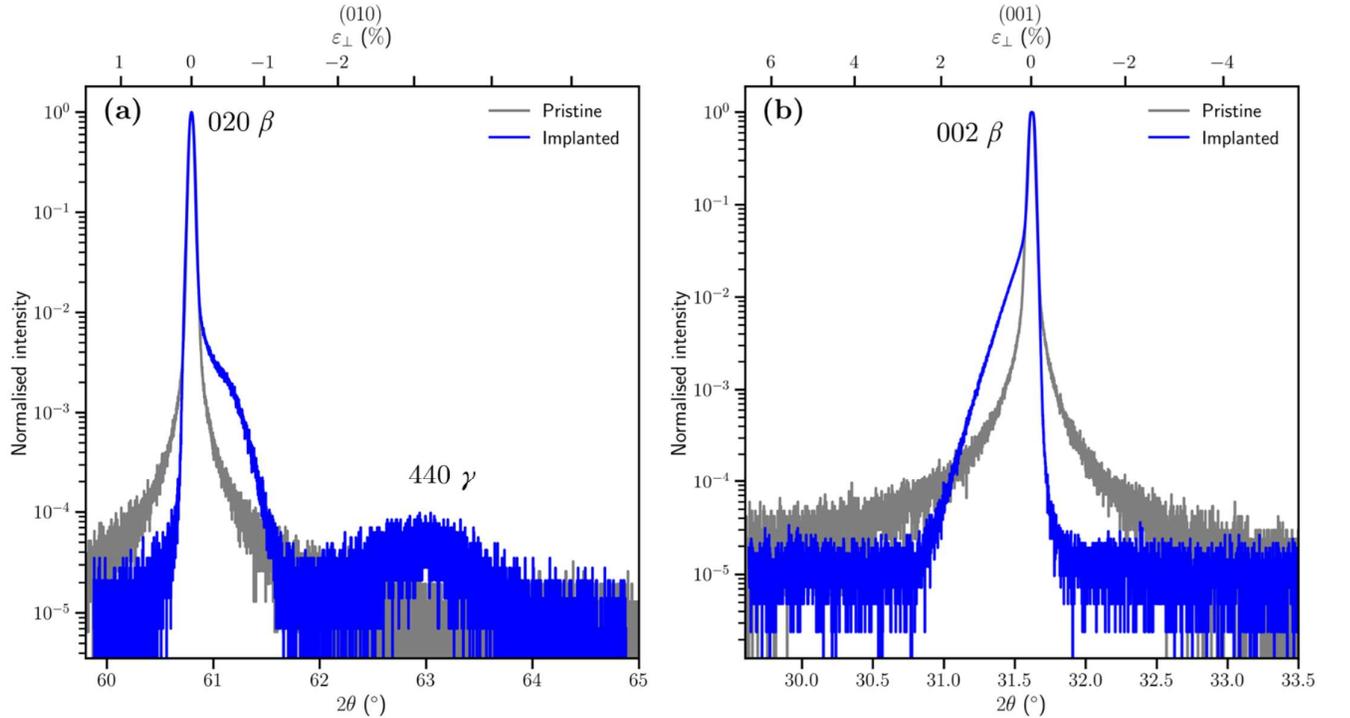

**Fig. S5** | Symmetric $2\theta - \omega$ scans about the 020 (a) and 002 (b) reflections of $\beta$-$Ga_2O_3$, obtained in (010) and (001)-oriented samples, respectively.

For the (010)-oriented samples, after the implantation, the peak associated with the 020 reflection of the $\beta$-phase shows a broadening towards larger values of $2\vartheta$ that correspond to the expected negative (i.e., compressive) strain along the [010] direction. Additionally, a clear broad peak is observed for values of $2\vartheta$ of about 63°, which is compatible with the 440 reflection of the $\gamma$-phase, revealing the $\beta$-to-$\gamma$ phase transition upon implantation, as reported previously by A. Azarov et al.[2]. Moreover, due to the symmetrical nature of the performed scans, this measurement also indicates that the (010) plane of the $\beta$-phase is parallel to the (110) plane of the $\gamma$-phase, suggesting a crystallographic relation between the two structures

For the (001)-oriented sample, after the implantation, the peak associated with the 004 reflection of the $\beta$-phase shows a broadening towards smaller values of $2\vartheta$ that correspond to a positive (i.e., tensile) strain along the direction perpendicular to the (001) planes. On the other hand, unlike for the (010) surface, no other peaks were observed. Given the observation of the $\gamma$-phase in the pole figures shown in the main paper, this may suggest that there is no low-index $\gamma$-phase plane that is parallel to the (001) plane of the $\beta$-phase, at least in this angular range. As shown in the main text, this is actually enforced by the crystallographic relation between the two polymorphs, and the $\gamma$-phase is indeed undetectable via symmetric $2\theta - \omega$ scans for the (001) surface orientation.

## S5. Simulated zone axis patterns

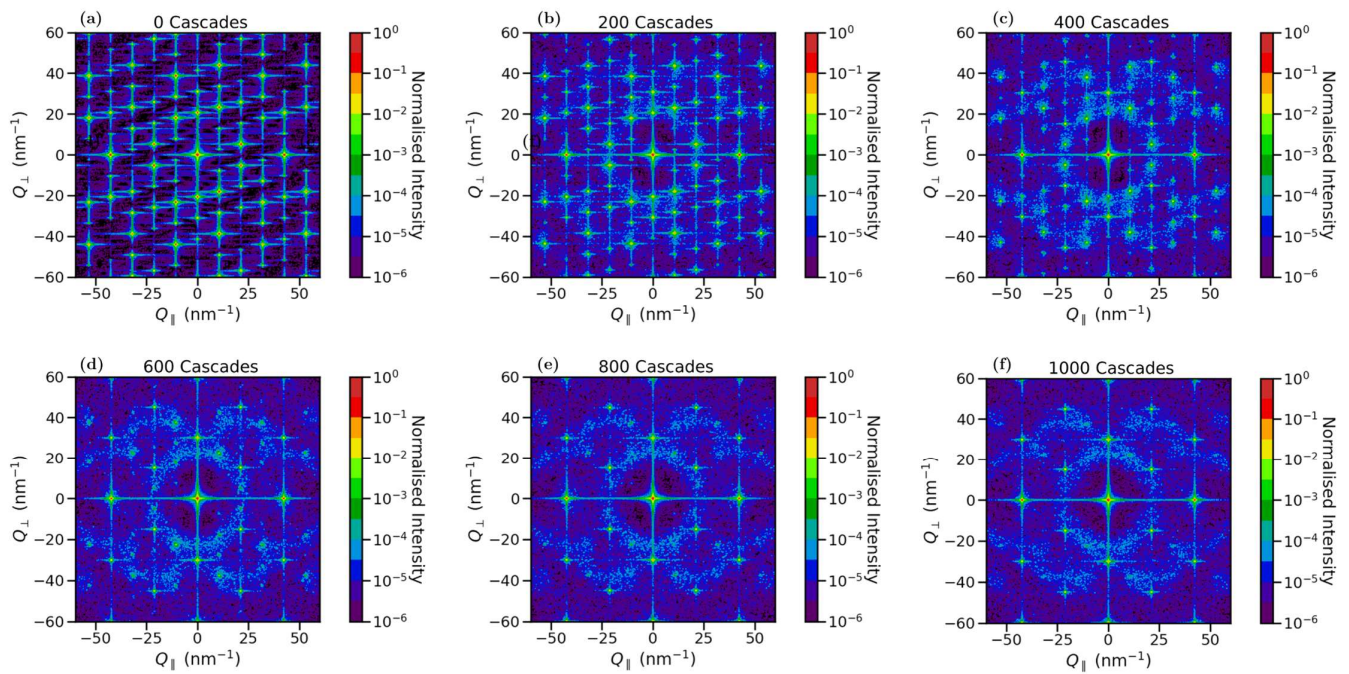

**Fig. S6** | Zone axis patterns obtained via MD simulations after (a) 0, (b) 200, (c) 400, (d) 600, (e) 800 and (f) 1000 overlapping cascades. These maps were obtained along the [010] zone axis of the original $\beta$-Ga$_2$O$_3$, which corresponds to the [110] zone axis of the induced $\gamma$-phase.

28/31

As described in detail in the end matter section of the main text, we performed Molecular Dynamics (MD) simulations with a large number of overlapping cascades (1000). This approach has been shown to accurately predict the formation of the $\gamma$-phase, as probed by, e.g., the radial distribution function[2]. In this work, we apply the method described in section 2.3 to simulate a large range of the reciprocal space of the sample, as shown in Fig. S6, obtained along the [010] zone axis. These simulated reciprocal space maps can be directly compared with experimental electron diffraction patterns, and show an excellent agreement. In Fig. S7, panels (a) and (f) of Fig. S7 are reproduced and superimposed on experimental Fast Fourier Transform (FFT) patterns applied to High-Angle Annular Dark-Field Scanning Transmission Electron microscopy (HAADF-STEM) images obtained along the [010] zone axis of pure $\beta$-$Ga_2O_3$ by J. García-Fernández et al.[3] and along the [110] zone axis of $\gamma$-$Ga_2O_3$ produced by ion implantation by A. Azarov et al.[2]. In conjunction with Fig. 4 on the main text, these results again confirm the proposed crystallographic relationship between the two phases: $(0\bar{1}0)_\beta \parallel (110)_\gamma$. Moreover, this demonstrates how MD simulations can be used to calculate reciprocal space quantities that are easier to compare with experimental results.

Additionally, Fig. S7 (e) shows the evolution of the integrated intensity in different regions of the reciprocal space, as indicated in Figs. S7 (a) and S7 (b). Remarkably, we can detect regions where a reflection associated with the $\beta$-phase is replaced by one associated with the $\gamma$-phase. For example, the $600$ and $\bar{2}04$ reflections are substituted by the $004$ and $440$ reflections, respectively. This is reflected by the increase or slight decrease in intensity with the number of overlapping cascades. A more interesting reflection is $802$ (which has a very low intensity), which is replaced by the $226$ reflection, with an intensity increase of almost two orders of magnitude. The intensity of the remaining marked regions simply decreases with the number of overlapping cascades. As such, this type of analysis is useful to assess phase transitions via MD as well as to design particular experiments to follow the phase transitions during irradiation.

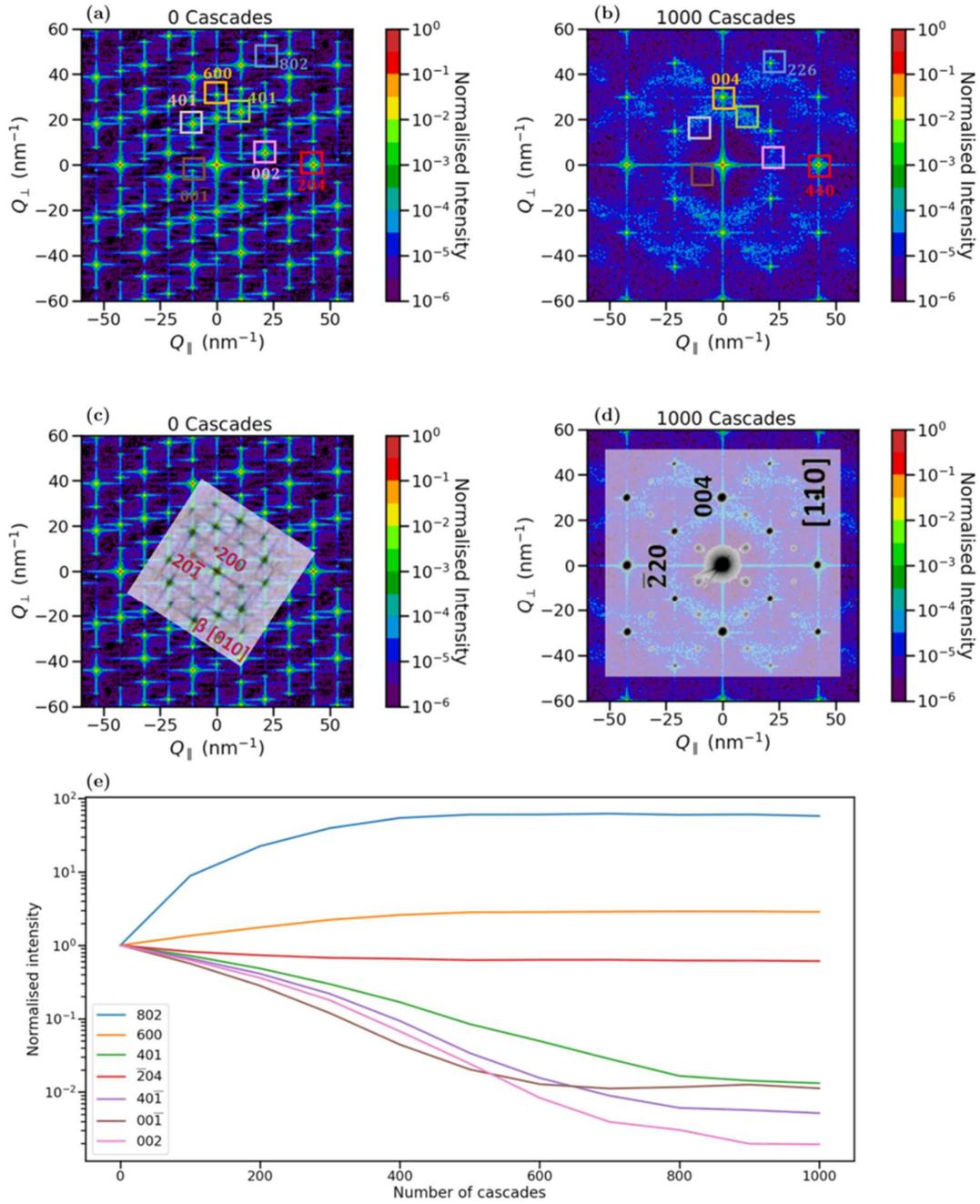

**Fig. S6** | Zone axis patterns obtained via MD simulations after (a) and (b) 1000 overlapping cascades, compared with experimental FFT patterns extracted from references (c) [3] and (d) [2] (which were rotated for comparison purposes). These maps were obtained along the [010] zone axis of the original $\beta$-$Ga_2O_3$, which corresponds to the [110] zone axis of the induced $\gamma$-phase. Panel (e) shows the evolution of the integrated intensity of the regions of the reciprocal space marked in (a) and (b) as a function of the number of overlapping cascades.